# A Framework for Ductility in Metallic Glasses

Sungwoo Sohn[1,✉,*], Naijia Liu[1,2,*], Geun Hee Yoo[1,3,*], Aya Ochiai[1], Jade Chen[1], Callie Levitt[1,4], Guannan Liu[1], Samuel Charles Schroers[1], Ethen Lund[1], Eun Soo Park[3], Jan Schroers[1,✉]

[1] Department of Mechanical Engineering and Materials Science, Yale University, New Haven, Connecticut 06511, USA
[2] Department of Materials Science and Engineering, Northwestern University, Evanston, IL, 60208, USA
[3] Department of Materials Science and Engineering, Seoul National University, Seoul 08826, Korea
[4] Department of Mechanical Engineering, University of Vermont, Burlington, Vermont 05405, USA
* These authors contributed equally to this work
✉ email: sungwoo.sohn@yale.edu, jan.schroers@yale.edu

## Abstract

The understanding and quantification of ductility in crystalline metals, which has led to their widespread and effective usage as a structural material, is lacking in metallic glasses (MGs). Here, we introduce such a framework for ductility. This very practical framework is based on a MGs' ability to support stable shear band growth, quantified in a stress gradient, $\nabla\sigma_{\mathrm{DB}}$, which we measure and calculate for a range of MGs. Whether a MG behaves ductile or brittle in an application is determined by the comparison between $\nabla\sigma_{\mathrm{DB}}$ and the applied stress field gradient, $\nabla\sigma_{\mathrm{app}}$; if $\nabla\sigma_{\mathrm{DB}} > \nabla\sigma_{\mathrm{app}}$, the MG will behave brittle, if $\nabla\sigma_{\mathrm{DB}} < \nabla\sigma_{\mathrm{app}}$, the MG will behave ductile, and $|\nabla\sigma_{\mathrm{app}}| - \nabla\sigma_{\mathrm{DB}}$ indicates how ductile. This framework can explain observed plastic properties of MGs and their apparent contradicting brittle and ductile characteristics. Looking forward, proposed framework provides the constitutive relation to quantitatively model their plastic behavior in any application, a requirement to use MGs as structural materials.

## Main text

### Introduction

Plastic deformation of crystalline metals is realized through dislocation motion which can result in a ductile response. In this manuscript, ductile is referred to as the room temperature ability to plastically deform and undergo macroscopic and permanent deformation in compression, tension and bending, which also reflects in high fracture toughness. The understanding and ability to quantify ductility in crystalline metals has laid the foundation for their widespread and effective usage as a structural material as it has allowed engineers to quantitatively model the plastic behavior in essentially any application [1]. Such ability does not exist for MGs.

In MGs, plastic deformation is strongly dependent on temperature and strain rate [2-4]. Deformation has been generally categorized into homogenous deformation occurring at high temperatures (relative to the glass transition temperature) and/or low strain rates, and into shear banding at low temperatures (including room temperature) and/or high strain rates [2]. MGs ubiquitously exhibit very high yield strengths and moderate modulus [5-7] (**Fig. 1**). Some MGs also display high fracture toughness [8-12]. Further, it has been observed that MGs during bending show bending ductility, when their thickness is below ~1 mm [13]. However, in contrast to the bending and fracture toughness behavior indicating a ductile behavior, all MGs, including the ones with high fracture toughness, are brittle under uniaxial tension. The exception are i.) MGs reduced to a small size, approximately below 100 nm due to the different scaling of involved energies [14] resulting in ductility [14-18] ii.) MG composites when length scales of the plastic zone size are matched with the spacing of the second phase [19, 20], iii.) MGs at very high potential energy where this somehow artificial state can lead to strain hardening [21], and iv.) in small samples when shear band propagation results in a stress drop within the sample, but only under conditions of a finite machine stiffness under displacement-controlled loading [22]. Under unconstrained compression, ductility has been reported even beyond these exceptions [8, 20, 23, 24]. However, careful Weibull analysis has suggested that testing conditions often depart from uniaxial loading due to imperfection in the test samples' geometry which has been suggested to give rise to ductility [25].

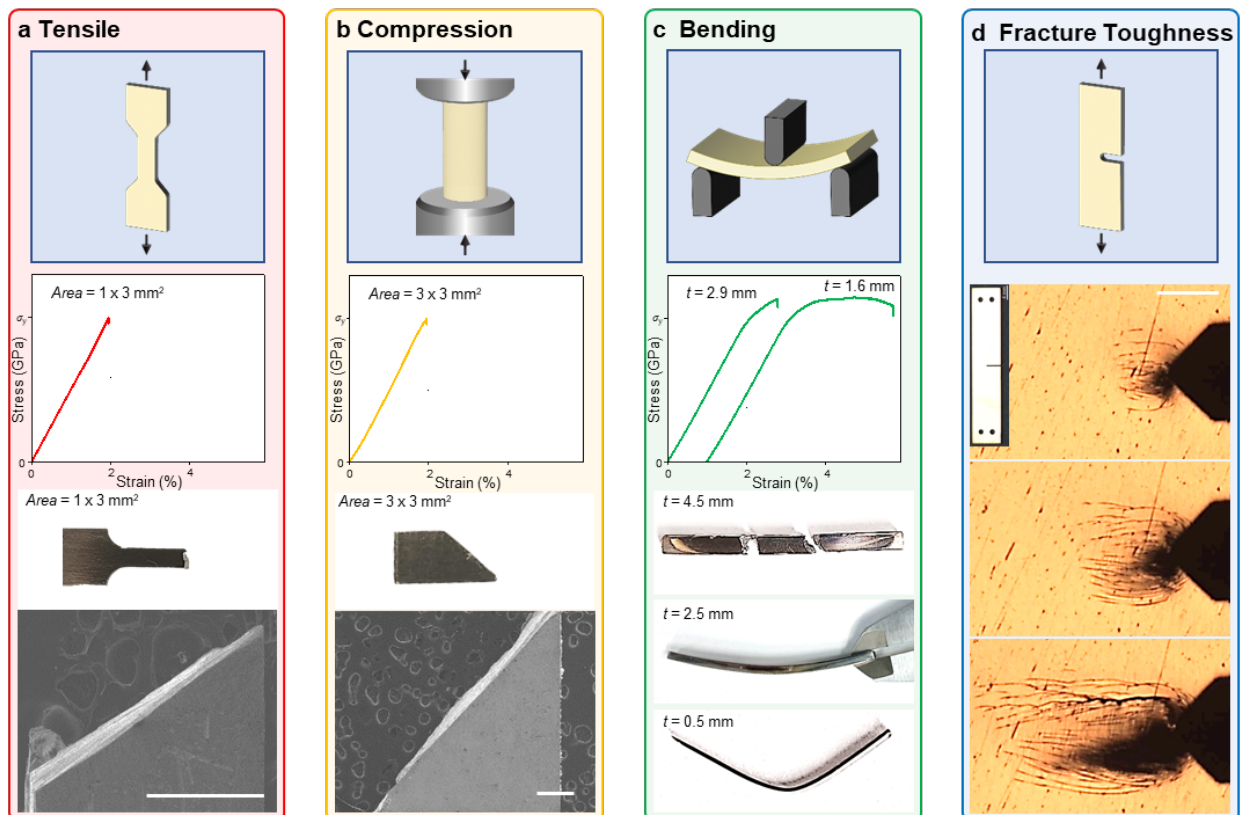

**Figure 1. Brittle and ductile characteristics of metallic glasses under standard testing conditions.**
**a-d.** Both brittle and ductile phenomena can be observed in MGs, here $Zr_{44}Ti_{11}Cu_{10}Ni_{10}Be_{25}$. **a, b.** Stress-strain curves and micrographs reveals brittle behavior in tensile and compression. Scale bars in (**a**, **b**) = 0.5 mm. **c, d.** Ductile behavior can be present in bending and fracture toughness. Scale bar in (**d**) = 0.2 mm.

The peculiar brittle and ductile characteristics of MGs (brittle in tension (**Fig. 1a**) and compression (**Fig. 1b**), ductile in bending (**Fig. 1c**) and fracture toughness (**Fig. 1d**), has been the focus of the MG community in the last 20 years, as it has prevented to draw conclusion about the MGs' plastic response in a general geometry, a requirement to use a material in structural applications. Further, even though ductile is a widely used term in the MG community, albeit, up to date, it cannot be quantified.

Heterogeneities and their resulting non-affine microscopic deformation have been suggested as the microscopic origin for ductile behavior [26-30]. Very similar in principle, various concepts for the description of heterogeneities such as atomic level stresses [31, 32], free volumes [33], soft-spots [34], pre-shear transformation zones (STZs) [35], minimum energy path in potential energy landscape [36], and low barrier crossings in the potential energy landscape [37] have been suggested. For their resulting non-affine microscopic deformation, STZs have been discussed [34, 37-44]. Even though a correlation between heterogeneities and ductility has been suggested [45], which has also been correlated with elastic constants [8, 41, 46, 47], no framework exists that allows to quantify and predict MGs' ductility. Further, it has been an open question how the plasticity that realizes ductility in MGs is achieved.

For a MG to exhibit room temperature ductility, shear bands must form and grow in a stable manner, where shear bands are able to grow without developing immediately into a crack and cause fracture. The growth of these shear bands is motivated by the applied stress field, $\sigma_{app}$. In the absence of strain hardening, but a general strain softening in MGs [18, 48-50], this motivation typically induces an unstable shear band growth. We argue that stable shear bands can in general only be achieved when the stress field drops faster than the deformation induced softening of the MG in the shear band. In other words, the gradient of the applied stress field, $\nabla\sigma_{app}$, must be larger than a specific value. This argument is key for the here developed framework for ductility in MGs. For a stable situation to occur, we introduce, provide a model, and measure the intrinsic stress gradient of a MG, $\nabla\sigma_{DB}$, when it can support stable shear band formation and hence ductility. $\nabla\sigma_{DB}$ is central to quantify the MGs' brittle or ductile behavior, and dependents only on the MG's chemistry and fictive temperature [41, 47].

$\nabla\sigma_{DB}$ quantifies the MG's stress gradient over which stable shear bands can form. Whether or not a MG behaves brittle or ductile is defined by how $\nabla\sigma_{DB}$ compares to the stress field gradient present in the sample, which is typically defined by the samples' geometry, $\nabla\sigma_{app}$. If $\nabla\sigma_{app} > \nabla\sigma_{DB}$, stable shear bands can be formed, the MG will behave ductile and $|\nabla\sigma_{app}| - \nabla\sigma_{DB}$ indicates how ductile. For $\nabla\sigma_{app} < \nabla\sigma_{DB}$, shear bands are unstable, and their formation is causing immediate fracture. In other words, a given MG is not intrinsically brittle or ductile but instead its brittle or ductile behavior depends on the environment, i.e., the stress field that is present in the MG sample. Thus, $\nabla\sigma_{DB}$ reveals the transition from forming stable to unstable shear bands. We argue that $\nabla\sigma_{DB}$ originates from the comparison of involved energies for a progressing shear band, the released elastic energy, and the resistance energies. The detailed discussion of the mechanism will be provided further below.

## Results

**Determination of the key factor for ductility; ductile-to-brittle transition stress gradient ($\nabla \sigma_{DB}$):**

Stress gradients, and therefore the possibility to determine $\nabla \sigma_{DB}$, are present in a wide range of geometries. For experimental convenience, we choose bending experiments to determine $\nabla \sigma_{DB}$ (**Fig. 2**). Specifically, we determine the maximum bending strain, $\epsilon = \frac{t}{2R}$, with $t$ as the beams' thickness, and $R$ as the radius of curvature, which we determine through mandrel bending or 4-point beam bending (**Fig. 2a**). The stress within the beam, $\sigma = \epsilon \cdot E$ varies linearly with the distance $y$ from the neutral axis. The stress gradient in the beam is given by $\frac{\sigma}{t/2}$ and the maximum value it can reach is $\frac{\sigma_{max}}{t/2} = \frac{\sigma_y}{t/2}$. To determine $\nabla \sigma_{DB}$ we fabricate and characterize beams of various thickness. The beams are deformed until fracture and subsequently characterized through microscopy to determine if shear bands have formed. The presence of multiple shear bands in the fractured beam is proof that stable shear bands have formed, as some of them have not developed into a crack. If multiple shear bands are observed, a thicker beam is fabricated, bended to fracture, and characterized to determine the number of shear bands. At some thickness, fracture occurs without formation of any stable shear bands; only one shear band forms that leads to fracture (7 mm for $Zr_{44}Ti_{11}Cu_{10}Ni_{10}Be_{25}$ in **Fig. 2b**). This thickness is the transition thickness from ductile to brittle behavior, $t_{DB}$, which gives $\nabla \sigma_{DB} = \frac{\sigma_y}{t_{DB}/2} = 570$ MPa mm$^{-1}$.

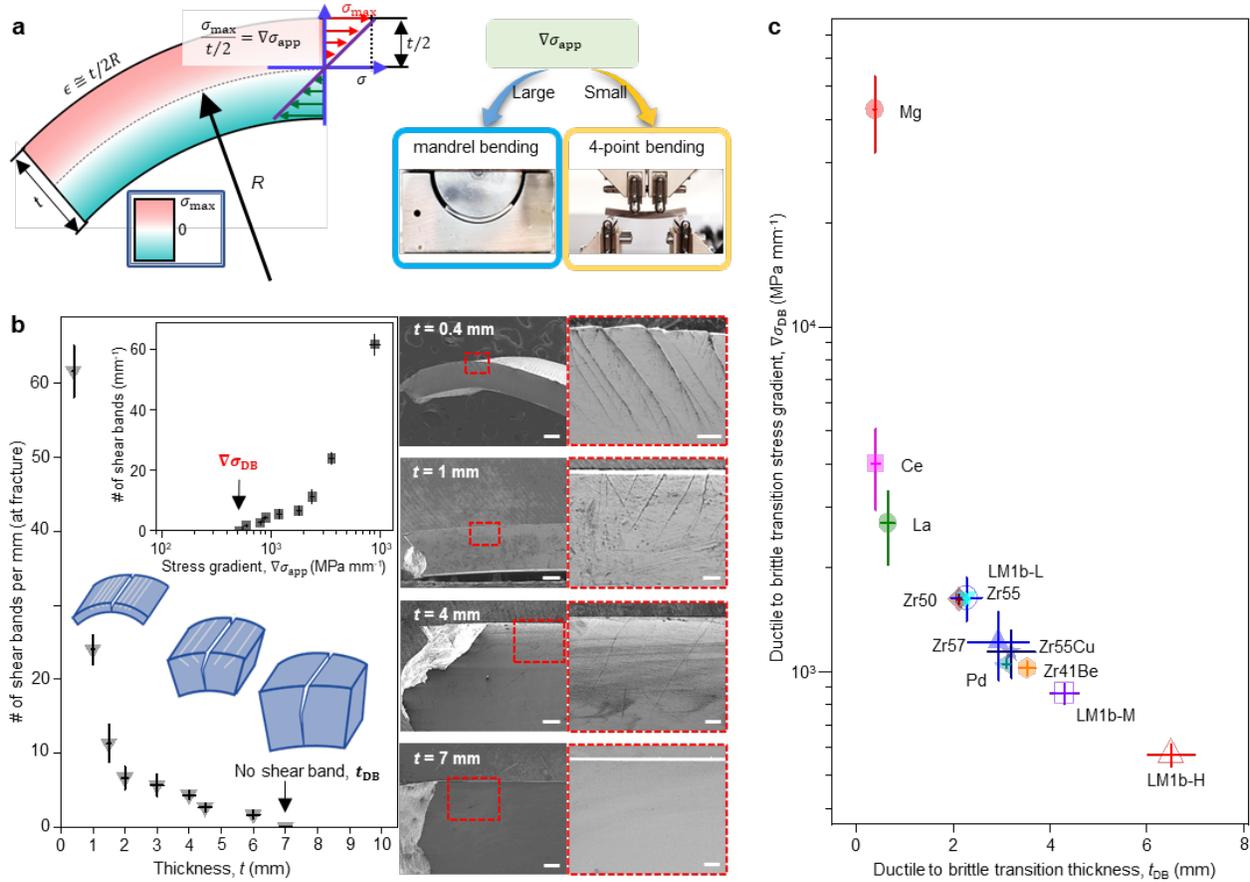

**Figure 2. Experimental determination of the ductile-to-brittle transition stress gradient.**
**a**. A schematic illustration of the stress gradient in a beam of thickness, $t$, which is bent to a radius, $R$. The maximum stress, $\sigma_{max}$, and the maximum stain, $\epsilon$, are present at the surface of the beam. The stress gradient is given by $\nabla\sigma_{app} = \frac{\sigma_{max}}{t/2}$ which can reach a maximum value at yielding of $\nabla\sigma_{app} = \frac{\sigma_y}{t/2}$ which we consider in the discussion of ductility. 4-point beam bending is used for characterization except for very thin samples < 0.4 mm where mandrel bending is used. **b**. A summary of the experiments carried out to determine $\nabla\sigma_{DB}$ for $Zr_{44}Ti_{11}Cu_{10}Ni_{10}Be_{25}$. Specimens with different thickness are bent to fracture and subsequently analyzed through microscopy for shear band formation. With increasing sample thickness, the number of stable shear bands decreases and at 7 mm thickness, no stable shear band can be observed in this alloy. We define this thickness as the transition thickness from brittle to ductile, $t_{DB}$, which gives for the ductile to brittle transition stress gradient, $\nabla\sigma_{DB} = \frac{\sigma_y}{t_{DB}/2}$. Inset shows the correlation between $\nabla\sigma_{DB}$ and the observed number of shear bands per mm. Scale bars in (**b**) $t$ of 0.4 mm = 200 & 50 µm, $t$ of 1 mm = 400 & 50 µm, $t$ of 4mm = 400 & 100 µm, $t$ of 7 mm = 400 & 100 µm. **c**. A summary of transition thickness, $t_{DB}$ and ductile to brittle transition stress gradient, $\nabla\sigma_{DB}$, of the here considered MGs. **Mg**: $Mg_{65}Cu_{25}Y_{10}$, **Ce**: $Ce_{60}Al_{20}Cu_{10}Ni_{10}$, **La**: $La_{55}Al_{25}Co_5Cu_{10}Ni_5$, **Zr57**: $Zr_{57}Cu_{15.4}Ni_{12.6}Al_{10}Nb_5$, **Pd**: $Pd_{43}Ni_{10}Cu_{27}P_{20}$, **Zr50**: $Zr_{50}Cu_{25}Al_{12.5}Ni_{12.5}$, **Zr55**: $Zr_{55}Cu_{22.5}Al_{11.25}Ni_{11.25}$, **Zr55Cu30**: $Zr_{55}Cu_{30}Al_{10}Ni_5$, **Zr41Be**: $Zr_{41.2}Ti_{13.8}Cu_{12.5}Ni_{10}Be_{22.5}$, **LM1b-L**: $Zr_{44}Ti_{11}Cu_{10}Ni_{10}Be_{25}$ with $T_f$ = 310°C, **LM1b-M**: $Zr_{44}Ti_{11}Cu_{10}Ni_{10}Be_{25}$ with $T_f$ = 340°C, **LM1b-H**: $Zr_{44}Ti_{11}Cu_{10}Ni_{10}Be_{25}$ with $T_f$ = 420°C.

**$\nabla\sigma_{DB}$ as an intrinsic property of a MG depending only on chemistry and fictive temperature:**

To represent the material class of MGs, we select alloys covering a wide range of fracture toughness, strength, and elastic modulus (**Fig. 2c** and **table 1**). We also consider alloys from the same alloy system and consider the same alloy at various fictive temperatures [41, 47].

**Table 1. Summary of the considered MGs.**
Chemical composition of MGs in at. %, $\sigma_y$: yield stress[10], $t_{DB}$: transition thickness from brittle to ductile, $K_Q$: fracture toughness[10], $\nabla\sigma_{DB}$: ductile to brittle transition stress gradient, $v$: Poisson's ratio, $G/B$: ratio of the shear modulus ($G$) to the bulk modulus ($B$), and fragility, $m$[51-53].

| Glass forming alloy (at.%) | $T_f$ (°C) | $\sigma_y$ (MPa) | $t_{DB}$ (mm) | $K_Q$ (MPa m$^{1/2}$) | $\nabla\sigma_{DB}$ (MPa mm$^{-1}$) | $v$ | $G/B$ | $m$ |
|---|---|---|---|---|---|---|---|---|
| Mg$_{65}$Cu$_{25}$Y$_{10}$ (Mg) | As-cast | 800 | 0.04 ± 0.03 | 2 | 43000 ± 10000 | 0.305 | 0.435 | 40 |
| Ce$_{60}$Al$_{20}$Cu$_{10}$Ni$_{10}$ (Ce) | As-cast | 800 | 0.4 ± 0.1 | 10 | 4000 ± 1000 | 0.313 | 0.422 | |
| La$_{55}$Al$_{25}$Co$_5$Cu$_{10}$Ni$_5$ (La) | 200 | 875 | 0.6 ± 0.2 | 27 | 2700 ± 700 | 0.340 | 0.358 | 37 |
| Zr$_{57}$Cu$_{15.4}$Ni$_{12.6}$Al$_{10}$Nb$_5$ (Zr57) | As-cast | 1785 | 3.0 ± 0.6 | 27 | 1200 ± 300 | 0.365 | 0.297 | 40 |
| Pd$_{43}$Ni$_{10}$Cu$_{27}$P$_{20}$ (Pd) | 360 | 1630 | 3.1 ± 0.1 | 40 | 1100 ± 40 | 0.399 | 0.215 | 65 |
| Zr$_{50}$Cu$_{25}$Al$_{12.5}$Ni$_{12.5}$ (Zr50) | 465 | 1800 | 2.1 ± 0.1 | 51 | 1700 ± 60 | 0.360 | 0.308 | |
| Zr$_{55}$Cu$_{30}$Al$_{10}$Ni$_5$ (Zr55Cu30) | As-cast | 1820 | 3.2 ± 0.5 | 58 | 1100 ± 180 | 0.369 | 0.287 | |
| Zr$_{55}$Cu$_{22.5}$Al$_{11.25}$Ni$_{11.25}$ (Zr55) | 470 | 1800 | 2.3 ± 0.1 | 64 | 1600 ± 50 | 0.373 | 0.276 | 68 |
| Zr$_{41.2}$Ti$_{13.8}$Cu$_{12.5}$Ni$_{10}$Be$_{22.5}$ (Zr41Be) | As-cast | 1800 | 3.5 ± 0.2 | 86 | 1000 ± 50 | 0.360 | 0.310 | 42 |
| Zr$_{44}$Ti$_{11}$Cu$_{10}$Ni$_{10}$Be$_{25}$ (LM1b-L) | 310 | 1860 | 2.3 ± 0.3 | 42 | 1600 ± 200 | 0.352 | 0.328 | 35, 43 |
| Zr$_{44}$Ti$_{11}$Cu$_{10}$Ni$_{10}$Be$_{25}$ (LM1b-M) | 340 | 1860 | 4.3 ± 0.3 | 75 | 870 ± 60 | 0.355 | 0.320 | |
| Zr$_{44}$Ti$_{11}$Cu$_{10}$Ni$_{10}$Be$_{25}$ (LM1b-H) | 420 | 1860 | 6.5 ± 0.5 | 109 | 570 ± 40 | 0.360 | 0.312 | |

The ductile to brittle transition stress gradient, $\nabla\sigma_{DB}$, varies broadly for different chemical compositions and fictive temperatures (**Fig. 2d** and **Table 1**). Among the considered alloys, the smallest $\nabla\sigma_{DB}$ of 570 ± 40 MPa mm$^{-1}$ ($t_{DB}$ ~ 6.5 mm) is present for Zr$_{44}$Ti$_{11}$Cu$_{10}$Ni$_{10}$Be$_{25}$ and the largest of over 43000 ± 10000 MPa mm$^{-1}$ ($t_{DB}$ ~ 0.04 mm) is measured for Mg$_{65}$Cu$_{25}$Y$_{10}$. The range of measured $\nabla\sigma_{DB}$ thus spans approximately two orders of magnitude. When considering alloys from one alloy system, $\nabla\sigma_{DB}$ also varies. For alloys from the Zr-Cu-Ni-Al system, $\nabla\sigma_{DB}$ = 1100 ± 180 MPa mm$^{-1}$ for Zr$_{55}$Cu$_{30}$Al$_{10}$Ni$_5$, $\nabla\sigma_{DB}$ = 1700 ± 60 MPa mm$^{-1}$ for Zr$_{50}$Cu$_{25}$Al$_{12.5}$Ni$_{12.5}$. The results further reveal that $\nabla\sigma_{DB}$ strongly depend on the fictive temperature, $T_f$. For Zr$_{44}$Ti$_{11}$Cu$_{10}$Ni$_{10}$Be$_{25}$ at a high $T_f$ = 420°C, 570 ± 40 MPa mm$^{-1}$ ($t_{DB}$ ~ 6.5 mm) and for a low $T_f$ = 310°C, 1600 ± 200 MPa mm$^{-1}$ ($t_{DB}$ ~ 2.3 mm).

**Mechanism for ductile and brittle behavior:**

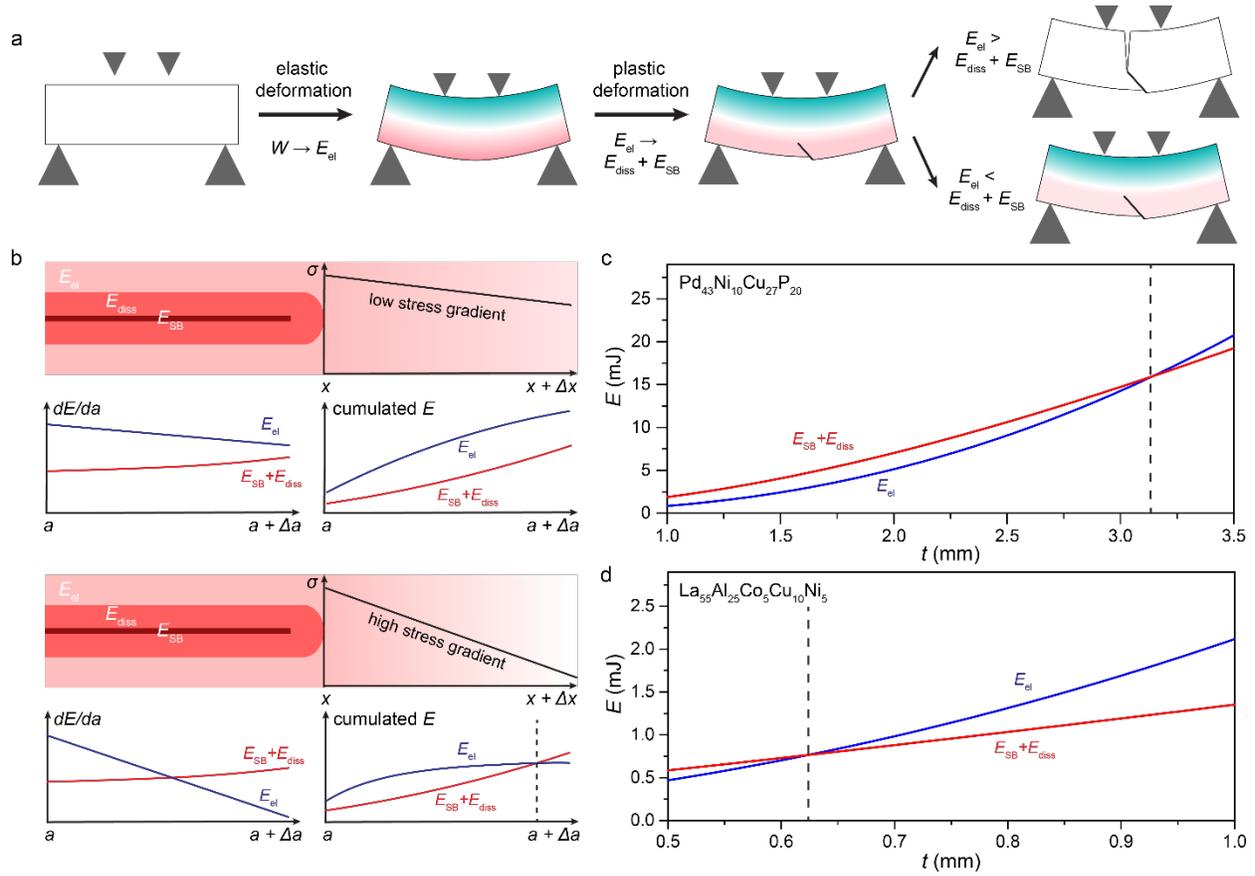

**Figure 3. Mechanistic origin for stable or unstable shear band growth in metallic glasses.**
**a.** The mechanical work to bend an MG specimen creates elastic energy. This elastic energy can be released through the shear displacement in a formed shear band. The released elastic energy, $E_{el}$, acts as the driving force for the progression of a shear band and the shear band energy and dissipated energy $E_{SB} + E_{diss}$ resist further progressing of the shear band. If for a forming shear band, the driving force drops below the resistance, $E_{el} < E_{SB} + E_{diss}$ it stabilized and results in ductile behavior. On the other hand, if $E_{el} > E_{SB} + E_{diss}$, the shear band is unstable, and causes fracture. **b.** The progressing behavior of a shear band is affected by the applied stress gradient, $d\sigma/dx$. The dependence of the rate of resist energy increase, $dE/dx = d(E_{SB} + E_{diss})/da$ (red line) on $d\sigma/dx$ is much weaker than the dependence of the driving force $dE/dx = dE_{elas}/da$ (blue line) on $d\sigma/dx$. This results in a smaller accumulated released energy with larger stress gradient, and the shear band is stable. In the stable case, a shear band will arrest when $E_{el}$ (blue) and the resistance energy $E_{SB} + E_{diss}$ (red) cross. **c.** For $Pd_{43}Ni_{10}Cu_{27}P_{20}$, our model yields a thickness where stable shear bands of length of 3.1 mm can be formed, compare to the experimentally determined $t_{DB} = 3.1$ mm. **d.** For $La_{55}Al_{25}Co_5Cu_{10}Ni_5$ as an example of a more brittle MG, our model yields a thickness where stable shear bands can form of 0.62 mm, compare to the experimentally determined $t_{DB} = 0.65$ mm.

We explain the brittle or ductile behavior of a MG by the ability to form stable shear bands, which originate from a net energy increase of the progressing shear band (**Fig. 3**). The balance of the net energy is discussed in the context of released elastic energy, shear band energy, and dissipated energy. The elastic energy, $E_{el}$, acts as a driving force for the progression of a shear band whereas the shear band energy, $E_{SB}$, and dissipated energy, $E_{diss}$, sit on the other side of the balance and hinder further growth of the shear band. In a loaded MG, the mechanical work from the loading instrument increases the stored elastic energy in the sample (**Fig. 3a** and detailed calculation in the Supplementary Material). This built-up elastic energy, $E_{el}$, can be released by the propagation of shear bands. This elastic energy generally forms the driving force for shear band propagation (**Fig. 3a**). The propagation of shear bands is typically resisted by two energy terms. One is the shear band energy, $E_{SB}$, the energy to form a shear band, which originates from the difference in the energy of the structure between the high $T_f$ shear band and the lower $T_f$ glass matrix and is typically on the order of $\gamma \sim 10$ Nm$^{-1}$ [54]. The other term is the generation of heat, essentially due to friction. This term can be estimated by the product of the local shear displacement, $\Delta u$, and friction, $\tau_f$. The friction term is a strong function of temperature due to the temperature dependence of the viscosity inside the shear band [55]. We refer to this associated energy as dissipated energy, $E_{diss}$. Since the shear rate within a shear band and the associated shear band propagation occurs by many orders of magnitude faster than thermal diffusion, $E_{diss}$ is contained in a thin shear affected zone as thermal energy. This thermal energy scales with the temperature increment and the size of the shear affected zone $w_s$. The dissipated energy can be quantified as $E_{diss} = aT_s\rho C_p w_s$ with $a$ as the shear band length, $T_s$ as the temperature inside the shear affected zone, and $\rho C_p$ as the volumetric heat capacity. If for a formed shear band $E_{el} < E_{SB} + E_{diss}$, the shear band will travel a final distance before it arrests. Such stable shear band enables plasticity and hence $E_{elas} < E_{SB} + E_{diss}$ indicates a ductile behavior. On the other hand, if $E_{el} > E_{SB} + E_{diss}$, a progressing shear band will not arrest, hence, once formed, the shear band will develop immediately into a crack and cause fracture, which is a brittle behavior (**Fig. 3a**).

To understand the stabilizing process of the applied stress gradient, we consider a shear band extending from $a$ to $a + \Delta a$. The increasing rate in the resist energies, $dE_{SB}/da = \gamma$ and $dE_{diss}/da = T_s\rho C_p w_s$ are both weak functions of the applied stress gradient. In contrast, the increasing rate of the driving force, $dE_{el}/da$, strongly depends on the applied stress gradient. For the given identical increment of shear displacement, $\Delta u$, a higher stress gradient results in a faster decrease of the local stress, $\sigma$, and thereby a faster-decreasing local density of the elastic energy, $\frac{1}{2}\sigma\epsilon$. Therefore, $dE_{el}/da$ decreases faster (**Fig. 3b**). This causes a faster decreasing $E_{el}$ than $E_{SB} + E_{diss}$, thereby stabilizing the shear band which allows for plasticity to yield a ductile behavior (**Fig. 3b**).

Based on these considerations, we develop a model to predict whether a stable shear band forms, hence ductility can be achieved. (**Fig. 3c-d**, and Supplementary Materials for details). We model the three energy terms to calculate the maximum length a shear band travels prior to transforming into a crack and compare the calculated values with the experimentally determined transition. As shown in **Fig. 3c**, theoretical derivation (see Supplementary Material for details) gives $E_{el} \propto (1/\nabla\sigma_{app})^2 \propto t^2$ and $E_{resist} = E_{SB} + E_{diss} \propto (1/\nabla\sigma_{app})^n \propto t^n$, (n < 2), as the applied stress gradient. As a consequence of their different scaling, the curves representing $E_{el}$ (blue) and $E_{resist}$ (red) cross at a specific sample thickness which defines the theoretical basis for $t_{DB}$. For a sample thickness smaller than $t_{DB}$, $E_{el} < E_{resist}$ and stable shear bands can form. For a sample thickness larger than $t_{DB}$, $E_{el} > E_{resist}$ and shear bands are unstable, prohibiting ductility. For example, for Pd$_{43}$Ni$_{10}$Cu$_{27}$P$_{20}$ with $E = 80$ GPa, $v = 0.3999$, $m = 65$ and $w_s = 10$ µm, the model predicts a transition from ductile to brittle behavior at ~3.1 mm (**Fig. 3c**), which compares well with the experimentally determined transition thickness of 3.1±0.1 mm (see **Table 1**).

The input parameters for the model calculating $t_{DB}$ are Youngs modulus, G/B, yield strength, fragility index ($m$), $T_g$, $\rho C_p$, shear band energy ($\gamma$), and $w_s$. We use in our model $w_s$ as a fitting parameter with values within the range of reported values [56-58]. For example, for a more brittle material, $La_{55}Al_{10}Co_5Cu_{10}Ni_5$, when assuming a smaller $w_s = 1$ μm, a similar $t_{DB}$ than experimentally determined of ~ 650 μm is calculated (**Fig. 3e**). Values for $w_s$ have been reported to vary from less than one micron to tens of micrometers for MGs [56-58]. A wide range of reported values may originate from the experimentally challenges in determine $w_s$. However, our model suggests that $w_s$ is MG specific and is generally narrower for brittle MGs. A MG specific $w_s$ is reasonable to assume with a variation from narrow to wide for brittle to ductile MGs. Brittle MGs are considered to exhibit a more dense and rigid structure [42, 46, 59-61]. Such structure cannot accommodate for a gradual and more spread-out deformation as the loosely packed and low shear resistant ductile structure. Novel experimental techniques are required to confirm our model prediction of an alloy specific shear affected zone thickness which appears to be key in determining brittle or ductile behavior of MGs.

In summary, by considering the energetics that promote and hinder shear band growth, the here introduced model can qualitatively describe when shear bands become stable or unstable. Thereby, it provides a mechanistic reasoning for the proposed framework for ductility: $\nabla \sigma_{DB} < \nabla \sigma_{app}$ => ductile and $\nabla \sigma_{DB} > \nabla \sigma_{app}$ => brittle. Further, for ductile samples where the first formed shear band arrests, the system will form more shear bands to further release the elastic energy and the strain level may keep increasing. When considering the increasing strain condition (and higher $\nabla \sigma_{app}$) of these later formed shear bands, the model can also estimate how much ductility can be achieved which is quantified in **Fig. 5**. It is important to mention that the purpose of the here presented model is simply to provide a mechanistic origin for the ductile or brittle behavior. For practical determination of a MGs behavior in any given application/geometry, use the here reported $\nabla \sigma_{DB}$ (or to determine it experimentally for MGs that are not considered here), and apply the here introduced framework.

**Discussion:**

We argue that the here presented framework with $\nabla \sigma_{DB}$ at its center and the comparison with $\nabla \sigma_{app}$ defines the ductile and brittle behavior of MGs. Hence, it must be able to explain the observed mechanical behavior of MGs in standard test geometries (**Fig. 4**). Under uniaxial tension $\nabla \sigma_{app} = 0$, hence $\nabla \sigma_{DB} > \nabla \sigma_{app}$. Therefore, following our framework, MGs will be generally brittle in tension (**Fig. 4a**). This is in line with today's finding that bulk metallic glasses (BMGs) do not exhibit ductility in tension, with the exceptions listed earlier. Under uniaxial compression also $\nabla \sigma_{app} = 0$, hence $\nabla \sigma_{DB} > \nabla \sigma_{app}$. Following our framework, MGs are therefore brittle in compression, with the exceptions listed earlier (**Fig. 4a**). This conclusion suggested by our framework appears to be contradicting the numerous reports on "ductility in compression" of MGs [8, 20, 23]. Here, we argue that the apparent "ductility in compression" originates from non-uniaxial conditions in the test samples, hence $\nabla \sigma_{app} \neq 0$. This situation can be quantified (**Fig. 4b**) by considering the degree of non-parallelism in the faces of the test samples, $\Delta$ resulting in $\nabla \sigma_{app} = E \frac{\Delta}{L}/t$ (see Supplementary Note 3 for details). With increasing $\Delta$, $\nabla \sigma_{app}$ increases and can reach a value larger than $\nabla\sigma_{DB}$ which would then allow stable shear band growth and hence, ductility.

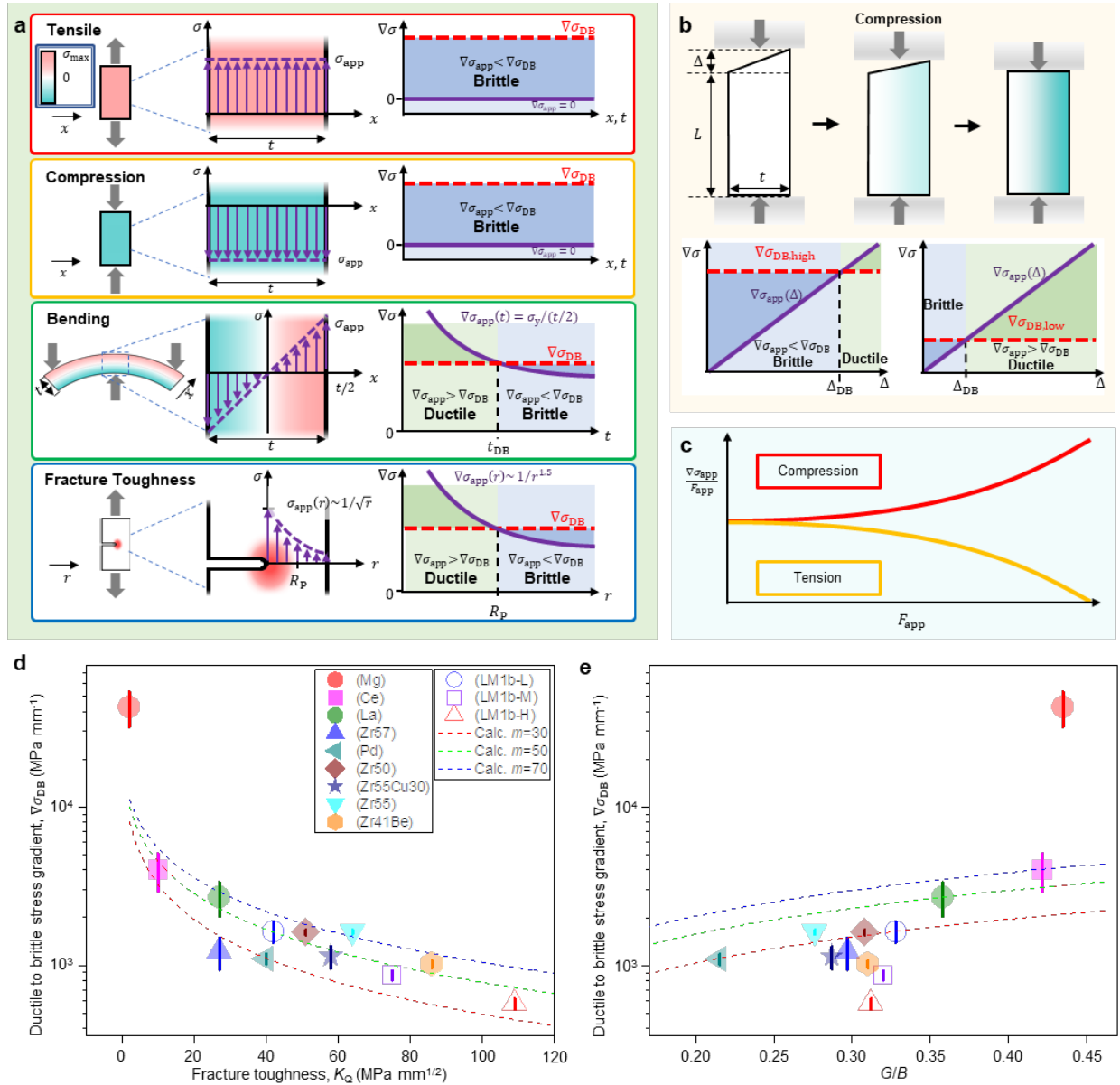

**Figure 4. Discussion of previous findings on mechanical properties of metallic glasses within the framework for ductility.**
a. Stress field distribution and stress gradient for tensile, compression, bending, and fracture toughness. $\nabla\sigma_{app}$ are compared with $\nabla\sigma_{DB}$. For uniaxial tension and compression $\nabla\sigma_{app} = 0$, wherefore for all MGs $\nabla\sigma_{app} < \nabla\sigma_{DB}$, hence no ductility can be realized in these geometries with MGs. In a bending geometry, $\nabla\sigma_{app}$ depends on the thickness and for $t < t_{DB}$, $\nabla\sigma_{app}$ becomes larger than $\nabla\sigma_{DB}$, hence a transition with sample thickness from brittle to ductile behavior with increasing sample thickness. In a fracture toughness test sample, the $1/\sqrt{r}$ stress field drop results in a $\nabla\sigma_{app}(r) \propto 1/r^{1.5}$. We argue here that $r(\nabla\sigma_{app} = \nabla\sigma_{DB})$ defines the plastic zone size, hence defining fracture toughness of MGs. For $r < r(\nabla\sigma_{app} = \nabla\sigma_{DB})$, $\nabla\sigma_{app} > \nabla\sigma_{DB}$, and shear bands progress in a stable manner. For $r > r(\nabla\sigma_{app} = \nabla\sigma_{DB})$, $\nabla\sigma_{app} < \nabla\sigma_{DB}$, shear bands become unstable and fracture occurs which terminates further formation of a plastic zone. **b.** Discussion of the previously determined ductility in tension and compression. Small specimen misalignment, $\Delta$, cause deviation from the homogenous stress field of an ideal uniaxial geometry and can lead, according to the here introduced framework, to ductility. **c.** As in tension the sample self-aligns, the stress field gradient is generally decreasing in tension. In contrast during compression, misalignments are further enhanced, wherefore with increasing applied load the stress gradients increase. **d-e.** Correlation of $\nabla\sigma_{DB}$ with other properties. **d.** $\nabla\sigma_{DB}$ vs. fracture toughness ($K_Q$). **e.** $\nabla\sigma_{DB}$ vs. the ratio of shear modulus ($G$) to bulk modulus ($B$), $G/B$. Blue, green, and red curves show calculated results with different fragility ($m$), $m = 30$, 50, and 70, respectively.

For example, assuming a test sample of 2 mm in length and 1 mm in diameter, $\Delta$ leading to ductility is 12 µm for $Zr_{44}Ti_{11}Cu_{10}Ni_{10}Be_{25}$, 26 µm for $Pd_{43}Ni_{10}Cu_{27}P_{20}$, 131 µm for $La_{55}Al_{25}Co_5Cu_{10}Ni_5$ and over 2000 µm for $Mg_{65}Cu_{25}Y_{10}$ (Supplementary Note 3). The required parallelism for $Zr_{44}Ti_{11}Cu_{10}Ni_{10}Be_{25}$ to result in $\nabla\sigma_{app}(\Delta) < \nabla\sigma_{DB}$ is difficult to realize through standard machining procedures, particularly in small samples. This explains the finding of "ductility in compression" where for MGs with small $\nabla\sigma_{DB}$, misalignment or non-parallel sample geometry can readily result in apparent ductility in compression. A small $\nabla\sigma_{DB}$ indicates even under small deviations from the uniaxial stress field a possible ductile response. For MGs with a large $\nabla\sigma_{DB}$, most notably $Mg_{65}Cu_{25}Y_{10}$, required stress field gradients to generate a ductile response are so high that they can be readily avoided in mechanical test samples. In fact, in most applications only a brittle response can be realized with this alloy. It is worth to discuss the discrepancy of apparent ductility in compression and tension of MGs (**Fig. 4c**). In compression, stress gradients normalized by the applied load increase with increasing loading. In tension, they decrease with increasing loading. This is because any misalignment in the sample is enhanced during compression whereas in tension the sample self-aligns. Hence, when the increasing load results in $\sigma = \sigma_y$, $\nabla\sigma_{app}$ has decreased substantially, making it very unlikely even in high non-parallelism in the test sample to observe ductility in tension. Hence, besides the exceptions discussed earlier, ductility in MG tensile test samples has not been reported.

In bending, ductility has been widely reported for MGs realized in beams which are thinner than ~ 1 mm [13, 47, 62]. Within introduced framework, and in contrast to the previous understanding, we argue that the thickness at which a MG exhibit bending ductility is an intrinsic property of the MG and does not ubiquitously occur at ~ 1mm. It varies by at least two orders of magnitude among MGs ranging from ~7 mm for $Zr_{44}Ti_{11}Cu_{10}Ni_{10}Be_{25}$ to less than 80 µm for $Mg_{65}Cu_{25}Y_{10}$ (**Fig. 2d** and **Table 1**). This is because $\nabla\sigma_{app}$ is inversely proportional to the beams' thickness, hence for the range of BMGs and their $\nabla\sigma_{DB}$ the condition for ductility, $\nabla\sigma_{app}(= \frac{\sigma_y}{t/2}) < \nabla\sigma_{DB}$ is fulfilled at different thicknesses (**Fig. 4a**).

Fracture toughness within the material class of MGs has been reported to vary from near-ideal brittle behavior to exceptionally tough [10]. The finding that MGs lack ductility under uniaxial testing but can exhibit exceptional high fracture toughness has been confusing and has remained a long-standing open question to this date. Introduced framework can now explain the confusing behavior and explain fracture toughness in MGs. Specifically, the corresponding size of the plastic zone, which controls fracture toughness is determined by the relative values of $\nabla\sigma_{app}$ and $\nabla\sigma_{DB}$ (**Fig. 4d**). The stress gradient in front of the plastic zone drops with the distance from the edge of the crack as $\nabla\sigma_{app}(r) \propto 1/r^{1.5}$ (**Fig. 4a**). Consequently, for small $r$, $\nabla\sigma_{app} > \nabla\sigma_{DB}$ which results in stable shear band progression. This continues, and the plastic zone is formed up to an $r$ where $\nabla\sigma_{app} < \nabla\sigma_{DB}$. Here, according to our framework, a brittle situation is present, which terminates further expansion of the plastic zone as the progressing shear bands develop into a crack and leads to fracture. Hence, the maximum size of the plastic zone, $R_p$, is set by the relative values of $\nabla\sigma_{app}$ and $\nabla\sigma_{DB}$ and $r(\nabla\sigma_{app} = \nabla\sigma_{DB}) = R_p$ defines the maximum size of the plastic zone, and hence, the fracture toughness of a MG. The ability of the framework to describe fracture toughness also reveals in the correlation between experimental data for fracture toughness and $\nabla\sigma_{DB}$ (**Fig. 4d**). Recent progress to fabricate BMG specific test samples that allow for intrinsic measurements of the fracture toughness [63] has also allowed to investigate correlations of fracture toughness with the ratio of shear modulus to bulk modulus, G/B [9, 10]. Even though some general correlations have been found, it has been concluded that the correlation holds for MGs from different alloy systems [10]. Similar here, when comparing $G/B$ with $\nabla\sigma_{DB}$ (**Fig. 4e** and Supplementary Materials), a much weaker correlation than between $\nabla\sigma_{DB}$ and fracture toughness is present (**Fig. 4d** and Supplementary Materials).

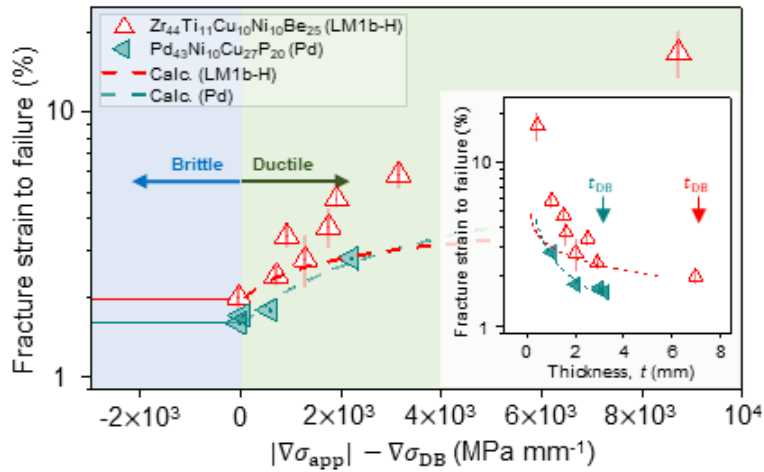

**Figure 5. Framework for ductility as a constitutive relation for the deformation of MGs.** $|\nabla\sigma_{app}| - \nabla\sigma_{DB}$ as a measure of ductility. As an example, fracture strain to failure ($\epsilon_{failure} = \epsilon_{elastic} + \epsilon_{plastic}$) as a function of $|\nabla\sigma_{app}| - \nabla\sigma_{DB}$ for $Pd_{43}Ni_{10}Cu_{27}P_{20}$ (Pd), and $Zr_{44}Ti_{11}Cu_{10}Ni_{10}Be_{25}$ with $T_f = 420°C$ (LM1b-H) are shown.

Going beyond the thus far considered binary case $\nabla\sigma_{app} > \nabla\sigma_{DB}$ => ductile and $\nabla\sigma_{app} < \nabla\sigma_{DB}$ => brittle, the introduced framework allows to quantify ductility of MGs in any given application (**Fig. 5** and Supplementary Materials). Let us consider the case when $\nabla\sigma_{app} > \nabla\sigma_{DB}$, e.g., in bending when the samples thickness is smaller than $t_{DB}$, Under this condition, stable shear band form and result in some plasticity. After the first shear band has formed, some local stress is released in the shear bands vicinity. The second shear band forms in a slightly higher stress field as the first shear band only released local stresses and the overall strain, hence stress has been increased. At some point, when the $n^{th}$ shear band forms, the stress level has reached such high level that $E_{el} > E_{SB} + E_{diss}$. At this point, the shear band becomes unstable and the sample fractures. The model does underrepresent the experimentally determined ductility as in the model it is not considered that for the case that many shear bands are formed, they can also release the far field stress, which results in a higher overall ductility. The specific conditions for this transition (fracture) to occur are experimentally determined for $Pd_{43}Ni_{10}Cu_{27}P_{20}$ and $Zr_{44}Ti_{11}Cu_{10}Ni_{10}Be_{25}$ (**Fig. 5**).

Figure 5 is a key outcome of introduced framework as it can be used as a constitutive relation to determine a MGs' deformation behavior in any geometry. This includes the specified geometries of mechanical test samples (**Fig. 4a**) but much broader, introduced framework allows for qualitative modeling of a MG's deformation behavior in any application (**Fig. 5**). This ability has been key for the usage of crystalline metals as structural materials where engineers and designers have been able to predict mechanical response in a specific application [1]. Through the introduced framework, this can now be done with MGs. Here, different inputs for the material properties and the geometry defining stress distribution are required for predicting deformation behavior for MGs than for crystalline metals. For a crystalline metal, the knowledge of the local stresses and its $\sigma(\varepsilon)$ acting as the constitutive equation allows for qualitative modeling of the elastic and plastic response. For MGs, qualitative modeling of their elastic and plastic response requires the knowledge of local stresses, local stress gradients, and the relationship of maximum strain to failure, $\varepsilon_f$ with $|\nabla\sigma_{app}| - \nabla\sigma_{DB}$ (**Fig. 5**). For quantitative modeling of the plastic response, $\varepsilon_f(|\nabla\sigma_{app}| - \nabla\sigma_{DB})$ provides the maximum local strain when the local stress level reaches $\sigma_y$. The actual deformation behavior to reach the maximum strain is MG specific and hence must be measured like here for $Pd_{43}Ni_{10}Cu_{27}P_{20}$ or $Zr_{44}Ti_{11}Cu_{10}Ni_{10}Be_{25}$ (**Fig. 5**). In the case that $\varepsilon_f(|\nabla\sigma_{app}| - \nabla\sigma_{DB})$ is unknown, $\varepsilon_F \times 100 = \frac{|\nabla\sigma_{app}| - \nabla\sigma_{DB}}{\nabla\sigma_{DB}}$ may provide a reasonable approximation.

In summary, we introduce a framework for ductility for MGs. This framework is based on the MGs' ability so support stable shear band growth, which is quantified in $\nabla\sigma_{DB}$, the minimum stress gradient over which growth of stable shear bands can take place. We measure $\nabla\sigma_{DB}$, which is a material property only depending on chemistry and fictive temperature, for a range of alloys to represent the material class of MGs, and construct a model, based on involved energies, that reveals the mechanistic leading to ductile or brittle behavior. If a MG behaves ductile or brittle in a given application is determined by the comparison between $\nabla\sigma_{DB}$ and the applied stress field, $\nabla\sigma_{app}$; If $\nabla\sigma_{DB} > \nabla\sigma_{app}$ the MG behaves brittle, if $\nabla\sigma_{DB} < \nabla\sigma_{app}$ the MG behaves ductile, and $|\nabla\sigma_{app}| - \nabla\sigma_{DB}$ indicates how ductile. This very practical framework can explain observed plastic mechanical properties of MGs and their apparent contradicting brittle and ductile characteristics. Looking forward, proposed framework provides the constitutive relation to quantitatively model their plastic behavior in any application, a requirement to use MGs as structural materials.

**Acknowledgements**
This research was supported by the Office of Naval Research under grant N00014-20-1-2200.